\documentstyle[a4,12pt,epsf]{article}
\begin{document}
\begin{titlepage}
\begin{flushright}
KUNS-1407 \\
HE(TH) 96/08 \\
Aichi-4/96 \\
\end{flushright}
\vspace{20pt}
\centerline{\huge Delay of Vehicle Motion}
\vspace{10pt}
\centerline{\huge in Traffic Dynamics}
        \vspace{40pt}
\centerline{\large Masako Bando\footnote
{e-mail address: bando@aichi-u.ac.jp},\ \
        Katsuya Hasebe\footnote{e-mail address: hasebe@aichi-u.ac.jp}}
\centerline{\small\it Physics Division, Aichi University,
                Miyoshi, Aichi 470-02, Japan}
        \vspace{20pt}
\centerline{\large Ken Nakanishi\footnote{e-mail address:
                nakanisi@gauge.scphys.kyoto-u.ac.jp}}
\centerline{\small\it Department of Physics, Kyoto University,
                Kyoto 606-01, Japan}
        \vspace{20pt}
\centerline{\large Akihiro Nakayama\footnote{e-mail address:
                g44153g@nucc.cc.nagoya-u.ac.jp}}
\centerline{\small\it Gifu Keizai University, Ohgaki, Gifu 503, Japan}

\vspace{60pt}

\begin{abstract}
We demonstrate that in Optimal Velocity Model (OVM) delay times of
vehicles coming from the dynamical equation of motion of OVM almost
explain the order of delay times observed in actual traffic flows
without introducing explicit delay times.  Delay times in various
cases are estimated: the case of a leader vehicle and its follower, a
queue of vehicles controlled by traffic lights and many-vehicle case
of highway traffic flow. The remarkable result is that in most of the
situation for which we can make a reasonable definition of a delay
time, the obtained delay time is of order 1 second.
\end{abstract}
\end{titlepage}

\section{ Introduction }
The history of traffic dynamics began early in 1950s. Two different
models were proposed: the one is car-following
model\cite{Gazis,Newell,Pipes}, which describe the motion of vehicles
by many-variable differential equations, and the other is the fluid
dynamical model\cite{Fluid}, which regard traffic flow something like
fluid.  These two models have been further investigated by many
authors. However either model did not succeed in explaining the
behavior of real traffic flow in some points. Especially those
traditional models can not attain an unified understanding the most
remarkable fact that traffic flow has two phases; one is free flow
with low car-density and high average speed and the other is congested
flow with high car-density and low average speed.

Recently, in these two models, there are found new mechanisms
which explain the existence of two phases. In car-following models,
a modified model was proposed by introducing optimal velocity\cite{AichiA}.
(We call this `Optimal Velocity Model' (OVM) hereafter.)
The equation of motion of OVM is given by
\begin{equation}
\ddot x_n(t) = a\{ V(\Delta x_n(t))-\dot x_n(t)\}\ \ \ \ n=1,2 \cdots N,
\label{eq:ovm0}
\end{equation}
where the notations are; car number $n$, time $t$, position of n-th 
car $x_n$ and its headway $\Delta x_n$.
A dot on the variable denotes differentiation with regard to time $t$,
sensitivity $a$ is a constant parameter. The essential difference from
traditional car-following models is the introduction of an optimal
velocity $V(\Delta x)$ of a vehicle, which value is a function
of headway distance. A driver reacts according to the difference
between the vehicle's velocity and the optimal velocity $V(\Delta x)$
and controls its velocity by accelerating (or decelerating) his
vehicle proportional to this velocity difference. 
The dynamical
equation of OVM has two different kind of solutions. One is a
homogeneous flow solution and the other is a congested flow
solution which consists of two distinct regions; congested
regions (high density) and free
regions (low density).
In OVM, if the density of vehicles is above some critical value
the traffic congestion occurs spontaneously, which can be
understood as a sort of phase transition from homogeneous flow state to
congested flow state\cite{AichiA,AichiB}. 

Also in fluid dynamical models, the following equation has been
proposed\cite{Kerner}.
\begin{equation}
\frac{\partial v}{\partial t}+v\frac{\partial v}{\partial x}=
\frac{V(\rho)-v}{\tau}-\frac{c_0^2}{\rho}\frac{\partial \rho}{\partial x}+
\frac{l^2}{\rho}\frac{\partial^2 v}{\partial x^2},
\label{eq:fluid0}
\end{equation}
where $c_0$, $\l$ and $\tau$ are constant parameters of the system and
density $\rho$ and velocity $v$ of vehicles are functions of location
$x$ and time $t$. They call the function $V(\rho)$ `safe velocity'. To
see the similarity of these two models, let us ignore second and third
term of rhs of Eq.(\ref{eq:fluid0}) and use $D/Dt=\partial/\partial t
+ v\partial/\partial x$; the differentiation acting on the variables
of individual vehicles.  The fluid dynamical equation
(\ref{eq:fluid0}) becomes
\begin{equation}
\frac{Dv}{Dt} = \frac{V(\rho)-v}{\tau} .
\end{equation}
This is quite similar to Eq.(\ref{eq:ovm0}) with optimal
velocity $V(\Delta x)$ and sensitivity $a$ replaced by `safe
velocity' $V(\rho)$ and parameter $1/\tau$,
respectively. The reason why these two models can explain the
formation of traffic congestion comes from the form of dynamical
equation and the introduction of `optimal' or `safe' velocity.

Contrary to this, the original equation of motion
of traditional car-following models\cite{Kometani}
\begin{equation}
\ddot x_n(t+\tau)=\lambda\{\dot x_{n-1}(t)-\dot x_n(t)\}\ ,
\label{eq:hist2}
\end{equation}
is essentially a first order differential equation without a delay
time $\tau$ of response. Because traffic flow 
governed by a first order differential equation is
always stable, the delay time $\tau$ plays a crucial role 
in describing the behavior of traffic flow.
The origin of the delay time $\tau$ have been thought to
be a physiological delay of response. In fact, it is well-known
that the motion of a vehicle accompanies some delay time
in response to the motion of its preceding vehicle.
There seem, however, questions to be discussed: what is the role
of the delay time $\tau$, or, whether or not the delay time $\tau$
is equal to the value of observed one.

Here we should remark that it is necessary to distinguish two different
types of definition concerning the notion of ``delay time''. There is
a time lag until the driver begins an action after being conscious of
a stimulus. It may also takes a finite time for a vehicle to change
its velocity after the operation of driver.  We define such
physiological and mechanical time lag as ``delay time of response''. 
This can be measured in principle although it may vary depending on
the type of the stimulus, individual driver and performance of
vehicle.  On the other hand, we can define ``delay time of vehicle
motion'' as a period from the time of velocity change of a vehicle to
that of following one. The delay time of vehicle motion can also be
estimated in observations of real traffic and/or numerical simulations
using traffic models. However in general case, it is difficult to find
a good definition of the delay time of motion of two successive
vehicles. Only in the restricted case in which two vehicles behave
quite similarly, for example,
\begin{equation}
v_n(t)\simeq v_{n+1}(t-T),
\label{eq:def_delay}
\end{equation}
we can define the delay time of motion as $T$.

In the car-following models, there are another two types of ``delay
time'' to be distinguished. One is the delay time $\tau$ explicitly
introduced as a parameter in the equation of motion (see
Eq.(\ref{eq:hist2})), which we call ``explicit delay time'' in this
paper. This may correspond to the delay time of response. The other is
the delay time emerging as a result of the dynamics of traffic flow,
which is quite different notion from the explicit delay time. This
will correspond to the delay time of vehicle motion stated above. 
However, this delay time includes not only the contribution from the
explicit delay time but also that from purely dynamical origin.  We
call the latter as ``effective delay time''.  Obviously, the effective
delay time is not zero even if we introduce no explicit delay time.

In this paper, we discuss the delay time of motion in OVM with 
no explicit delay time. Therefore, because the
resulting delay time is equal to the effective delay time,
we can investigate the delay time from dynamical origin.
This quantity may be compared with observed delay
times of vehicle motion in various cases. It will be clear that 
the effective delay time obtained by our procedure is enough to
explain the delay times observed in actual situations of traffic flow.

In section 2, we define the effective delay time of vehicle motion in
OVM in terms of the vehicle motions of a leader and its follower and
make an analytical study within linear approximation. Then we carry
out numerical
simulations to obtain the effective delay time in several cases.
We show the results in actual traffic
situations: the effective delay times under traffic rights
 (section 3) and those in
uniform traffic flow and in congested flow (section 4).  Discussions
and further prospects are given in section 5.

\section{Delay Time for Leader and Follower Case} 
First let us make an definition of effective delay time of vehicle
motion.  Consider a pair of vehicles, a leader and its follower. This
pair of vehicles may be either separate two vehicles or any pair of
vehicles picked up from a series of vehicles in highways or from a
queue waiting to start under traffic lights.

When a leader moves with its velocity $v(t)$, and its follower
replicates the motion of the leader with some delay time $T$ (then the
follower's velocity is given by $v(t-T)$), we can define the delay time of
vehicle-motion as $T$. It must be remarked that we do not define the
delay time as the delay of motion of the follower with respect to its
position but its velocity replication.

Let the positions of a leader and its follower be $y(t)$ and $x(t)$.
In this case Eq.(\ref{eq:ovm0}) is written as
\begin{equation}
 \ddot{x}(t) = a\{ V(y(t) - x(t))-\dot{x}(t)\}. 
\label{eq:leader_follower_eq}
\end{equation}
Uniform motion is described by
\begin{equation}
y_0(t)=V(b)t + b,\ \ x_0(t)=V(b)t,
\label{eq:const_motion}
\end{equation}
where $b$ is headway and $V(b)$ is a constant velocity.
To investigate the response of the follower vehicle to the leader
vehicle, we introduce a small perturbation $\lambda(t)$ and its
response $\xi(t)$:
\begin{equation}
y(t)=y_0(t) + \lambda (t), \ x(t)=x_0(t) + \xi(t).
\label{eq:per_to_const_motion}
\end{equation}
Inserting Eq.(\ref{eq:per_to_const_motion})
into Eq.(\ref{eq:leader_follower_eq}) and taking a linear 
approximation, we get
\begin{equation}
\ddot{\xi}(t) + a\dot{\xi}(t) + af \xi(t) = af \lambda(t) , 
\label{eq:leader_follower_leq}
\end{equation}
where $f = V'(b)$. This is just equivalent to well-known equation of
motion for forced oscillation with damping term caused by friction.

In order to find a solution, we first write $\lambda(t)$ 
by a Fourier expansion
\begin{equation}
\lambda(t) = \int \tilde {\lambda}(\omega)e^{i\omega t} d\omega \ .
\end{equation}
For the Fourier component $\lambda_0e^{i\omega t}$,
the solution of Eq.(\ref{eq:leader_follower_leq}) is given by
\begin{equation}
\xi(t) = \frac{\lambda_0}{1 + i\omega/f - \omega^2/af}
\ e^{i\omega t}.
\label{eq:E8}
\end{equation}
This is rewritten as
\begin{equation}
\xi(t) = |\eta|\lambda_0 e^{i\omega (t - T)},
\label{eq:E9}
\end{equation}
where
\begin{eqnarray}
|\eta|^2&=&\frac{|\xi|^2}{\lambda_0^2}
=\frac{(af)^2}{(af-\omega^2)^2+(a\omega)^2},
\label{eq:amp_LA}\\
T &=& \frac{1}{\omega}\tan^{-1}\frac{a\omega}{af-\omega^2}.
\label{eq:delay_LA}
\end{eqnarray}
When $f<a/2$, the amplitude $|\eta|$ is monotonically damping function of
$\omega$. On the other hand, when $f>a/2$, $|\eta|$ takes its maximum
at $\omega=\omega_0$;
\begin{equation}
\omega_0^2 = a(f -a/2),
\label{eq:omega0}
\end{equation}
so we call this $\omega_0$ as ``enhanced mode'' 
(see Fig.\ref{fig:eta-delay}(a)). Note that $f=a/2$ is
the critical point for the instability condition of homogeneous flow
as we found in the previous paper \cite{AichiA}. 
Eq.(\ref{eq:omega0}) shows that the enhanced mode
 $\omega_0\ (\neq0)$ exists so
far as the instability condition $f > a/2$ is satisfied. 

Let us examine some characteristic cases. For 
$f<a/2$ where low frequency modes dominate $|\omega|\ll a,\ f$, we have
\begin{equation}
|\eta| \sim 1, \ T \sim \frac{1}{f}.
\label{eq:xiandt}
\end{equation}
In this case the response $\xi(t)$ to the perturbation $\lambda(t)$
becomes
\begin{equation}
\xi(t)= \int \tilde {\lambda}(\omega)e^{i\omega (t - T)} d\omega \ 
= \lambda(t - T),
\end{equation}
which leads
\begin{equation}
\dot x(t)=V(b)+\dot\xi(t)=V(b)+\dot\lambda(t - T)=\dot y(t-T) .
\label{eq:repli}
\end{equation}
Thus for sufficiently slow perturbation, the delay time of vehicle motion
becomes $T$ in Eq.(\ref{eq:xiandt}), which is approximately 
the inverse of derivative
of the optimal velocity function at corresponding headway.

\begin{figure}[htb]
\hspace*{-0.6cm}
\epsfxsize=7.5cm
\epsfbox{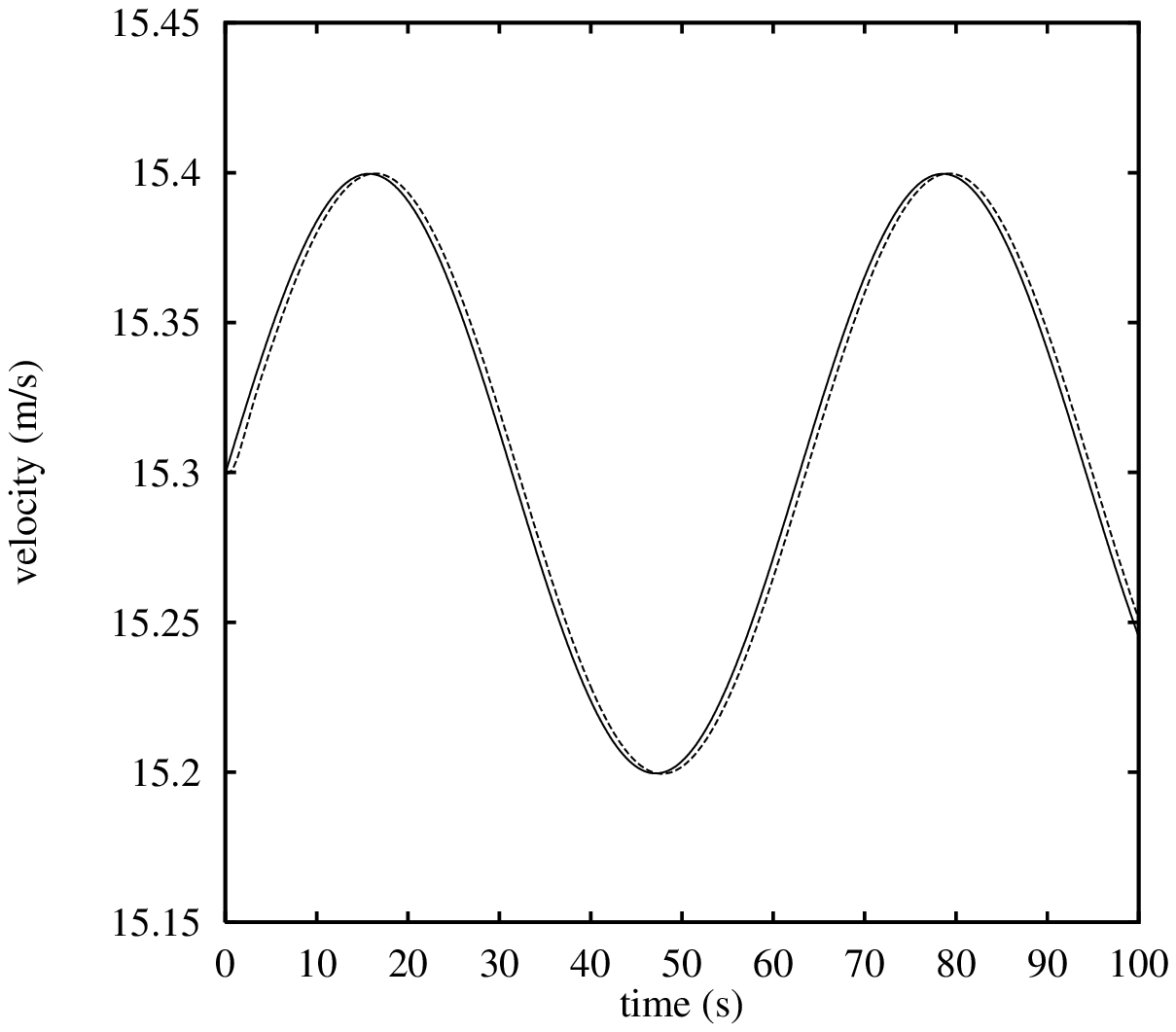}
\epsfxsize=7.5cm
\epsfbox{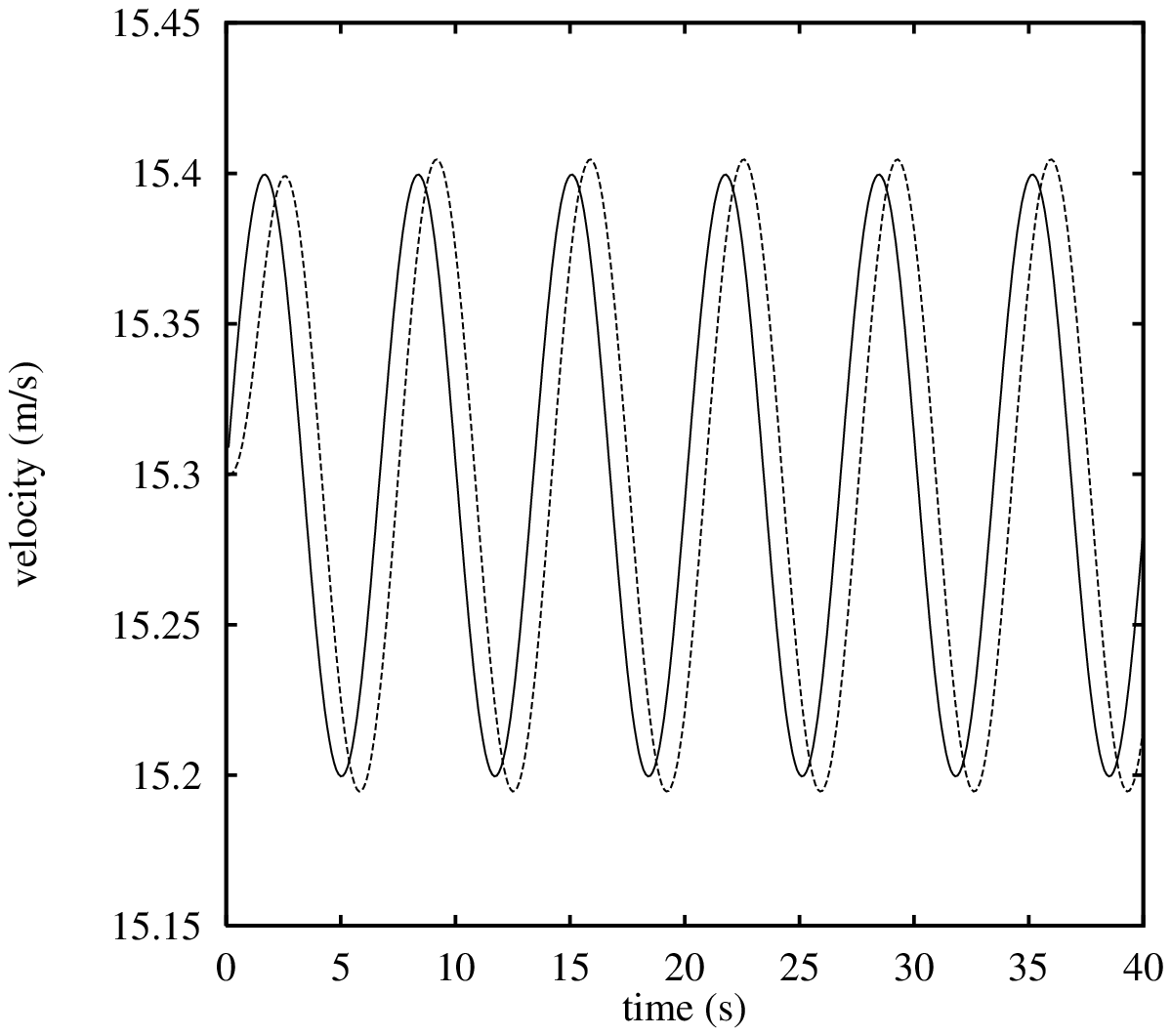}\\
\hspace*{-0.6cm}
\epsfxsize=7.5cm
\epsfbox{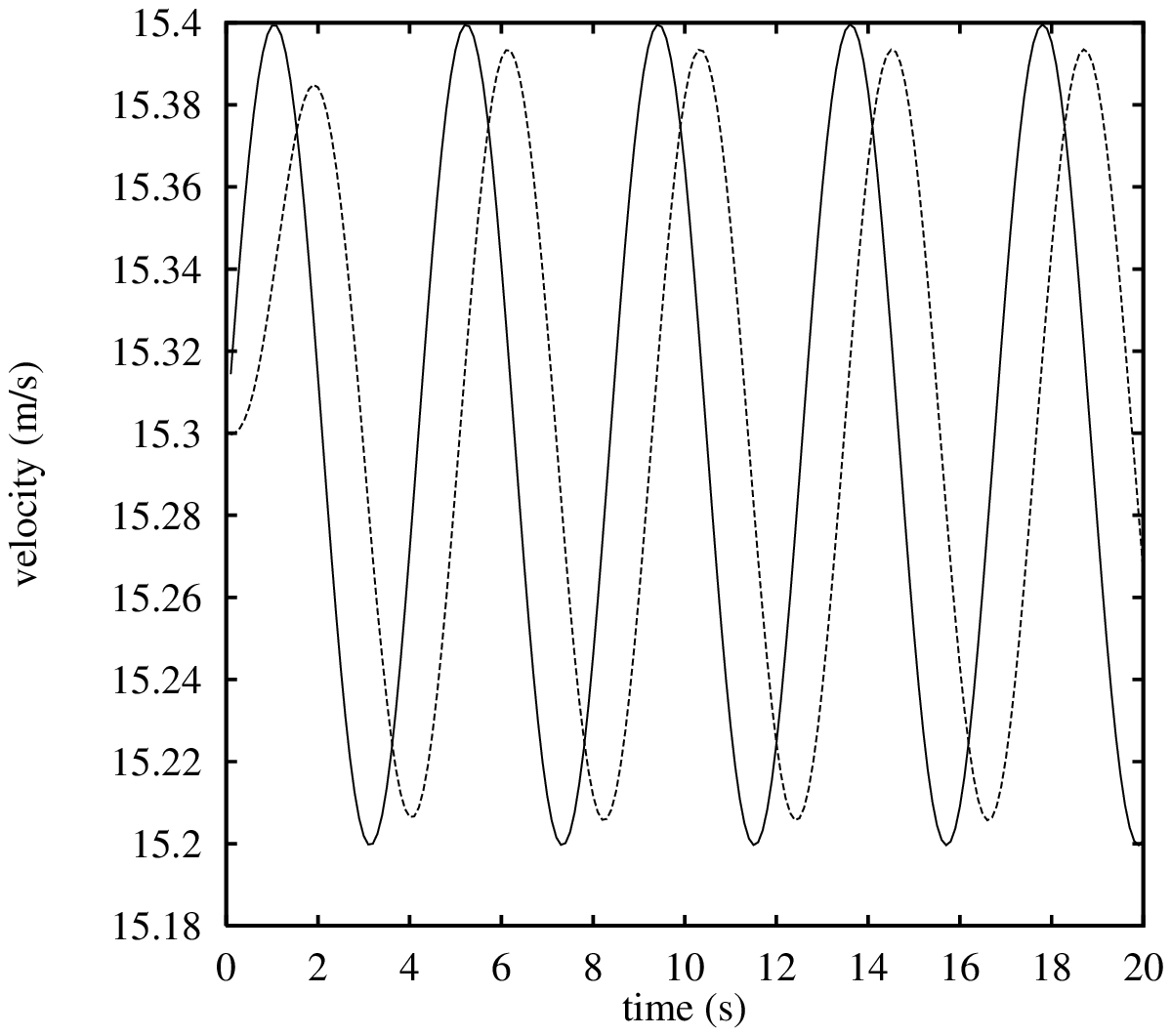}
\caption{Numerical results for the motion of leader and follower,
where $\Delta x=$25~m and $a=2.0~{\rm s}^{-1}$.
Frequencies of leader's motion are (a) $\omega=0.1~{\rm s}^{-1}$,
(b)~$\omega=\omega_0$ and (c) $\omega=1.5~{\rm s}^{-1}$.}
\label{fig:LF1}
\end{figure}

\begin{figure}[htb]
\hspace*{-0.3cm}
\epsfxsize=7cm
\epsfbox{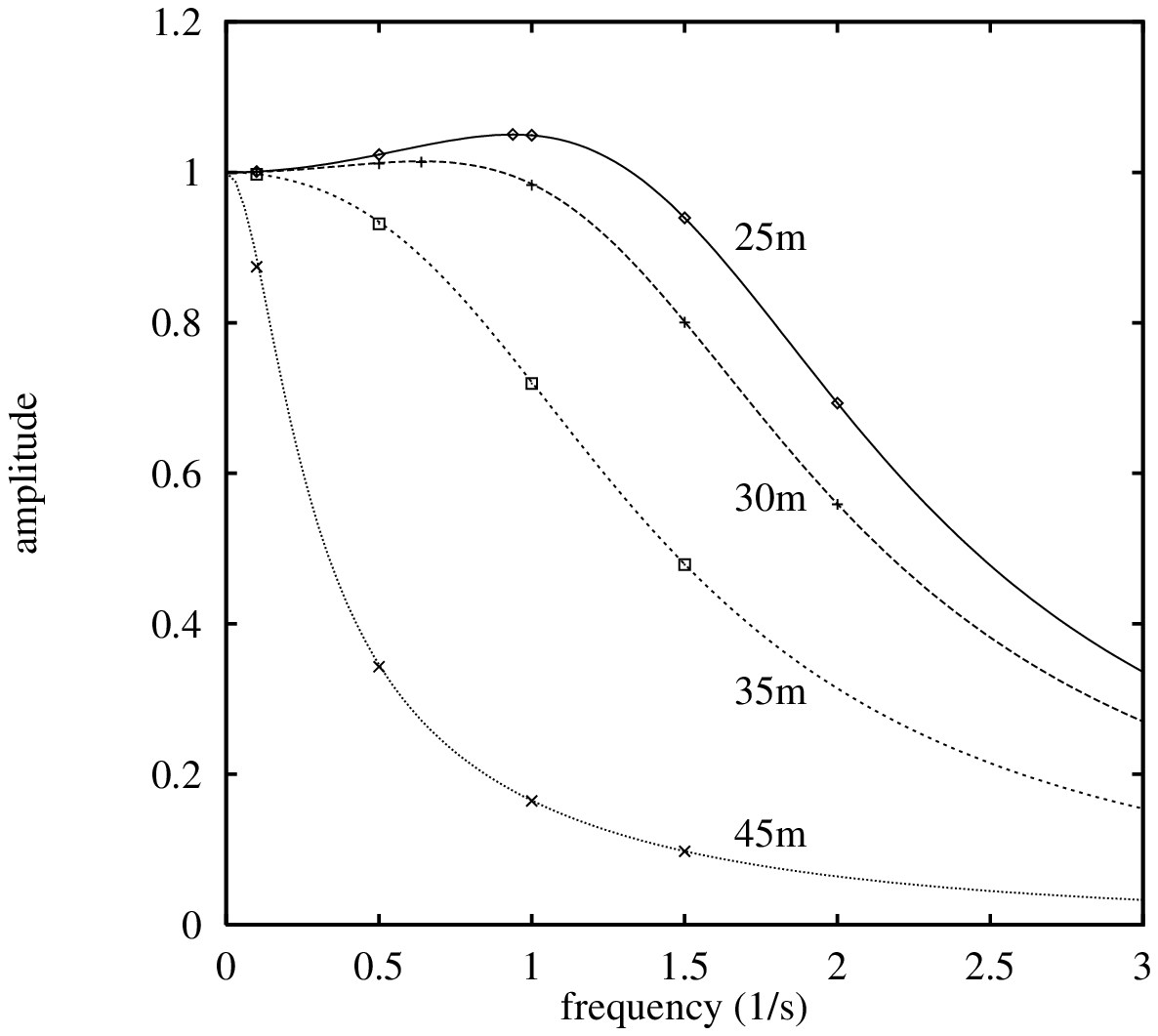}
\hspace*{0.3cm}
\epsfxsize=7cm
\epsfbox{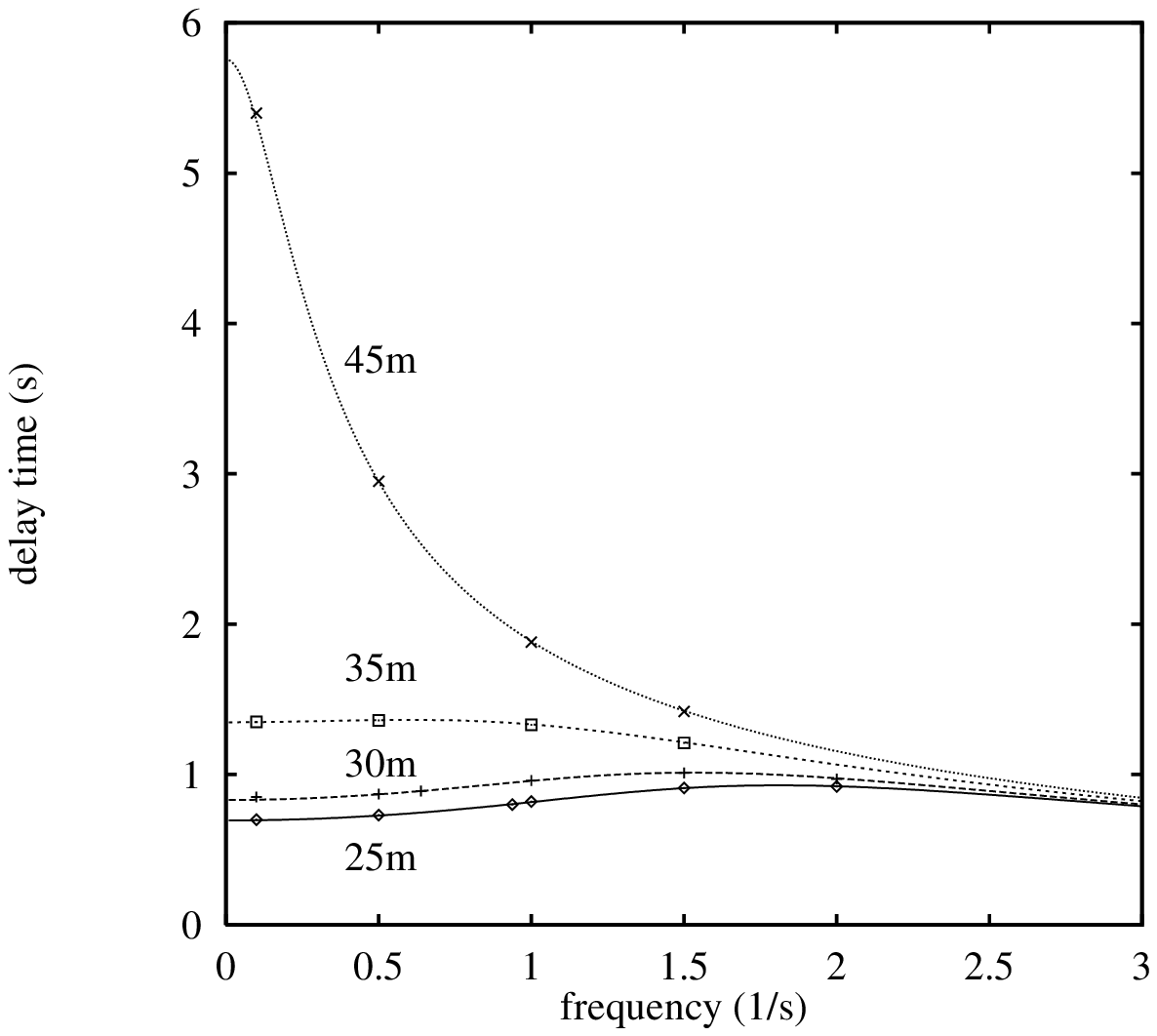}
\caption{Each curve shows the behavior of (a) $|\eta|$
(Eq.(\protect{\ref{eq:amp_LA}}))
and (b) $T$ (Eq.(\protect{\ref{eq:delay_LA}}))
for $b=25,\ 30,\ 35,\ 45$~m with $a=2.0~{\rm s}^{-1}$.
Plotted marks on the curves show numerical results.}
\label{fig:eta-delay}
\end{figure}

In the other case $f>a/2$, the amplitude $|\eta|$ takes maximum
value at $\omega=\omega_0$. Then we have
\begin{equation}
T_{\rm enhanced} = \frac{1}{\omega_0}\tan^{-1}\frac{2\omega_0}a,
\label{eq:enhanced}
\end{equation}
which indicates that the delay time $T$ for this enhanced mode depends
on the sensitivity $a$ in contrast to the previous case
Eq.(\ref{eq:xiandt}). One can easily confirm that $T$ tends to $1/f$ when
$a$ is close to its critical value $2f$. We should remark that an
exact replication indicated in Eq.(\ref{eq:repli}) is not realized in
this case because $|\eta|$ is not always equal to 1 in
Eq.(\ref{eq:E9}).  (See also Figs.\ref{fig:LF1} and the discussions
below Eq.(\ref{eq:lam0}).)

Now let us see the results of numerical simulations and compare them
with those of our analytical consideration. Here and hereafter we use
the following form of the optimal velocity function \cite{AichiC},
\begin{eqnarray}
V(\Delta x)= 16.8\left[\tanh{0.0860\left(\Delta x - 25\right)} +
    0.913\right] &({\rm for}\ \Delta x > 7{\rm m}) \\
=0 \qquad\qquad\qquad\qquad\qquad\qquad\qquad\quad
 &({\rm for}\ \Delta x < 7{\rm m})
\end{eqnarray}
whose parameters are determined from the 
car-following experiment on Chuo Motorway \cite{Oba,KoshiIO}:
the inflection point is $(\Delta x,\dot x)=(25~{\rm m},55~{\rm km/h})$,
the maximal velocity is $V_{\rm max}=115~{\rm km/h}$ and the minimal
headway is $\Delta x_{\rm min}=7.0~{\rm m}$, which includes the length
of the vehicle (5~m) which was used in the experiment.

In numerical simulations, we prepare a pair of vehicles with their
unperturbed motions of Eq.(\ref{eq:const_motion}) with headway $b$. If
the leader changes its motion by $\lambda(t)$ then the function
$\xi(t)$ can be obtained by the numerical simulation from which the
delay time can be read off. We choose $\lambda(t)$ of the leaders
motion as
\begin{equation}
y(t) = V(b)t + b - \lambda_0 \cos{\omega t}, \quad
\dot y(t) = V(b) + \lambda_0 \omega \sin{\omega t} \quad
{\rm for}\ t\ge0
\label{eq:lam0}
\end{equation}
with various $\omega$ setting $\omega\lambda_0=0.1$~m/s, which means
that $|\dot y(t) - V(b)| \le 0.1$~m/s. 

As illustrations, we show the behaviors of $\lambda(t)$ and its
response $\xi(t)$ for $a=2.0~{\rm s}^{-1}$, $b=25$~m (therefore
$\omega_0 =0.938~{\rm s}^{-1}$). Figures \ref{fig:LF1} are the cases
for $\omega = 0.1,\ 0.938,\ 1.5~{\rm s}^{-1}$.
To find the values $T$, we first rescale $\dot\xi(t)$ 
and then translate it so as to coincide the curve $\dot\lambda(t)$
in Figures \ref{fig:LF1}.
The value of $|\eta|$ is this scale factor.
The numerical simulations have performed for $b=25,\ 30,\ 35,\ 45$~m.
In Figures \ref{fig:eta-delay}, 
we show the numerical values for $|\eta|$ and $T$.
The corresponding analytical results
Eqs.(\ref{eq:amp_LA}) and (\ref{eq:delay_LA}) are also shown in these
figures. In both figures, the analytical and numerical results
agree quite well.

\section{Delay Time under Traffic Lights}

The delay time of vehicle motion is clearly recognized in motions of a
series of vehicles under traffic lights. Consider the situation in
which every vehicle waits until a red light changes to green. The
initial conditions are as follows;
\begin{eqnarray}
&&\Delta x_1(0)=\infty,\quad\quad\,  \dot x_1(0)=0\cr
&&\Delta x_n(0)=7({\rm m}),\quad \dot x_n(0)=0\qquad(n=2,3,\dots)
\end{eqnarray}
If the light changes to green, the top vehicle will first accelerates
to start, followed by the succeeding vehicles according to the
equations of motion. 
\begin{figure}[htb]
\hspace*{-0.6cm}
\epsfxsize=7.5cm
\epsfbox{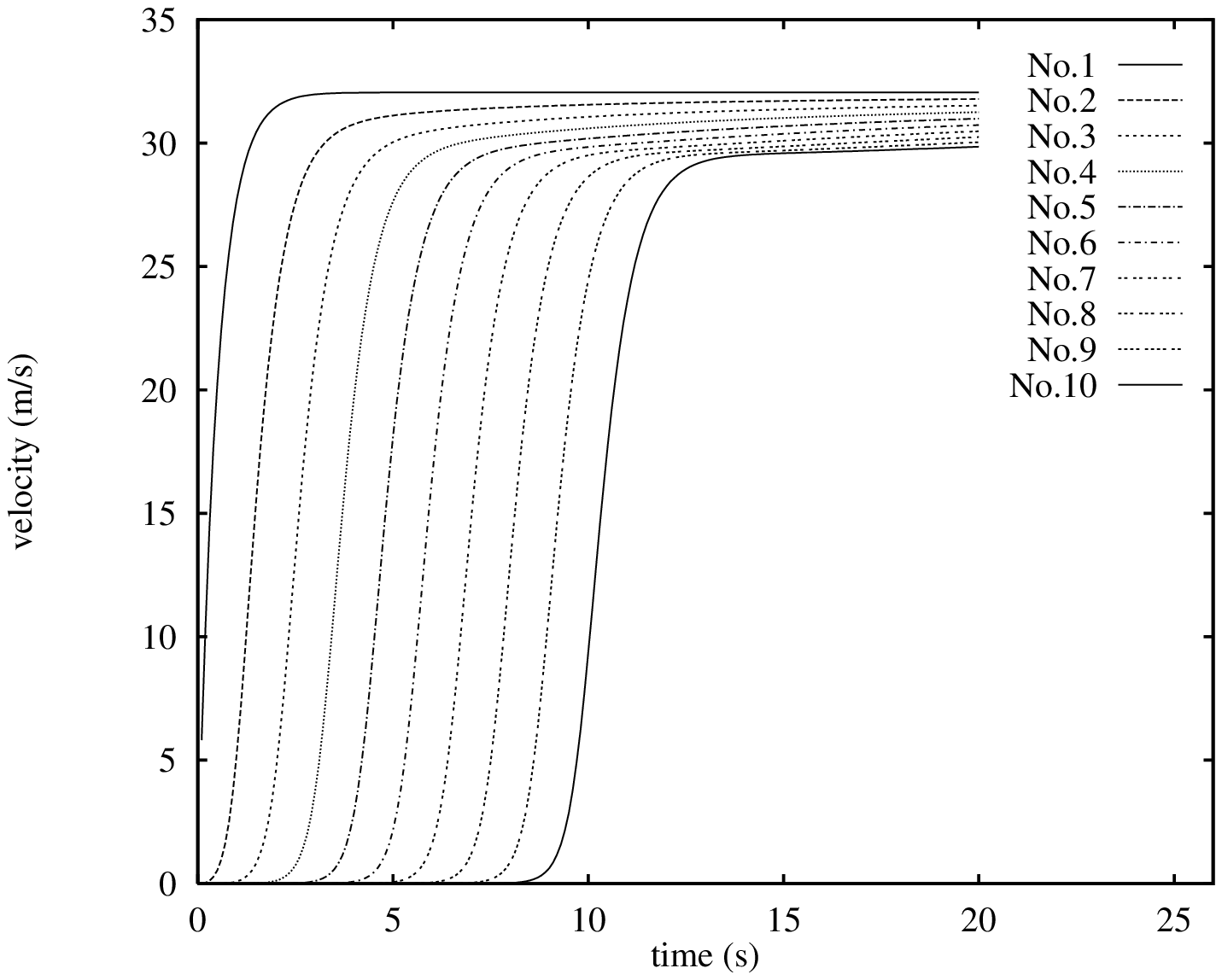}
\epsfxsize=7.5cm
\epsfbox{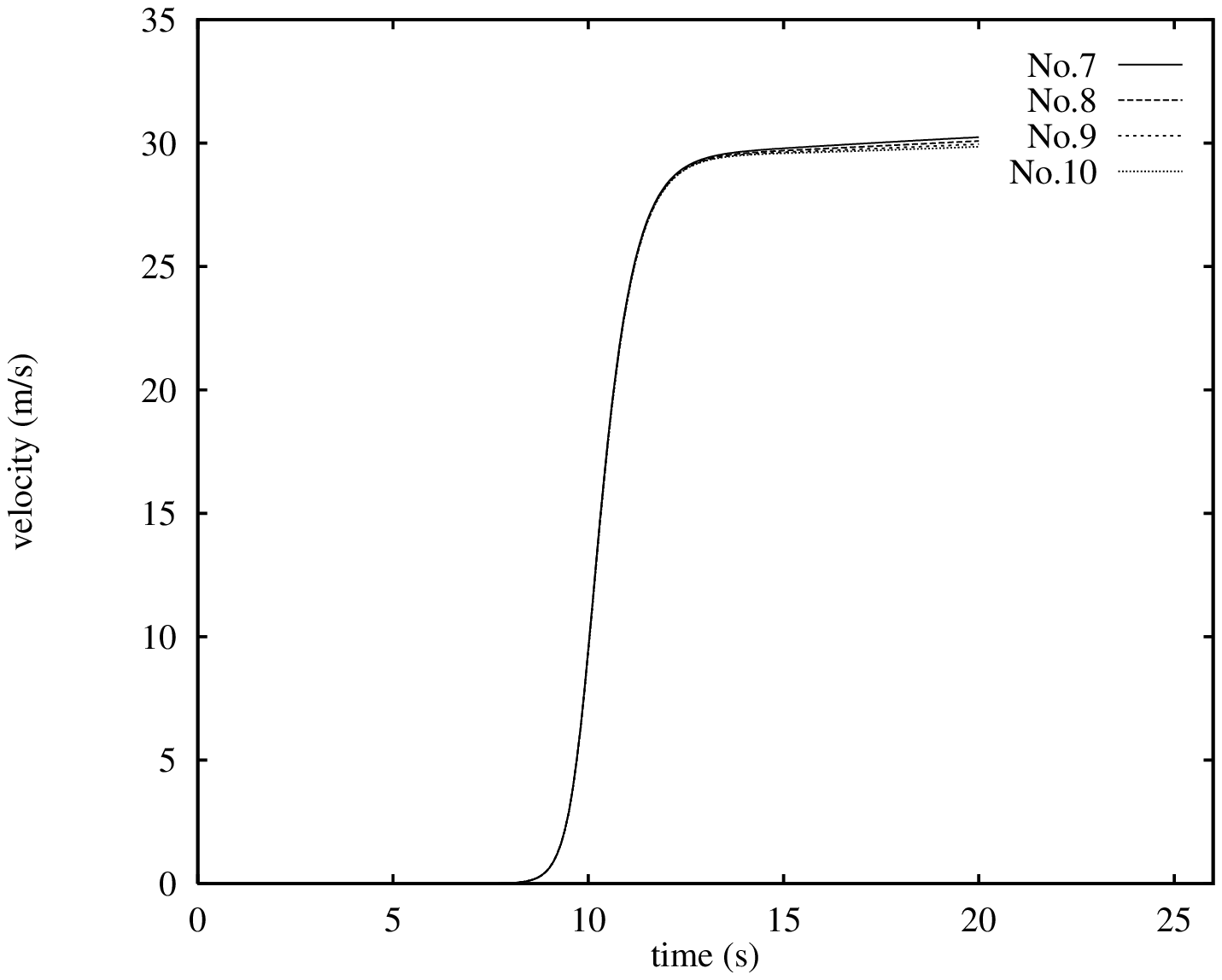}
\caption{(a) Behavior of velocities of first ten vehicles
under traffic lights with $a=2.0~{\rm s}^{-1}$.
(b) Figure of shifted curves $\dot x_n(t-(n-10)T)$ with $T=1.10$~s.}
\label{signal.2.0}
\end{figure}
\begin{figure}[htb]
\hspace*{-0.6cm}
\epsfxsize=7.5cm
\epsfbox{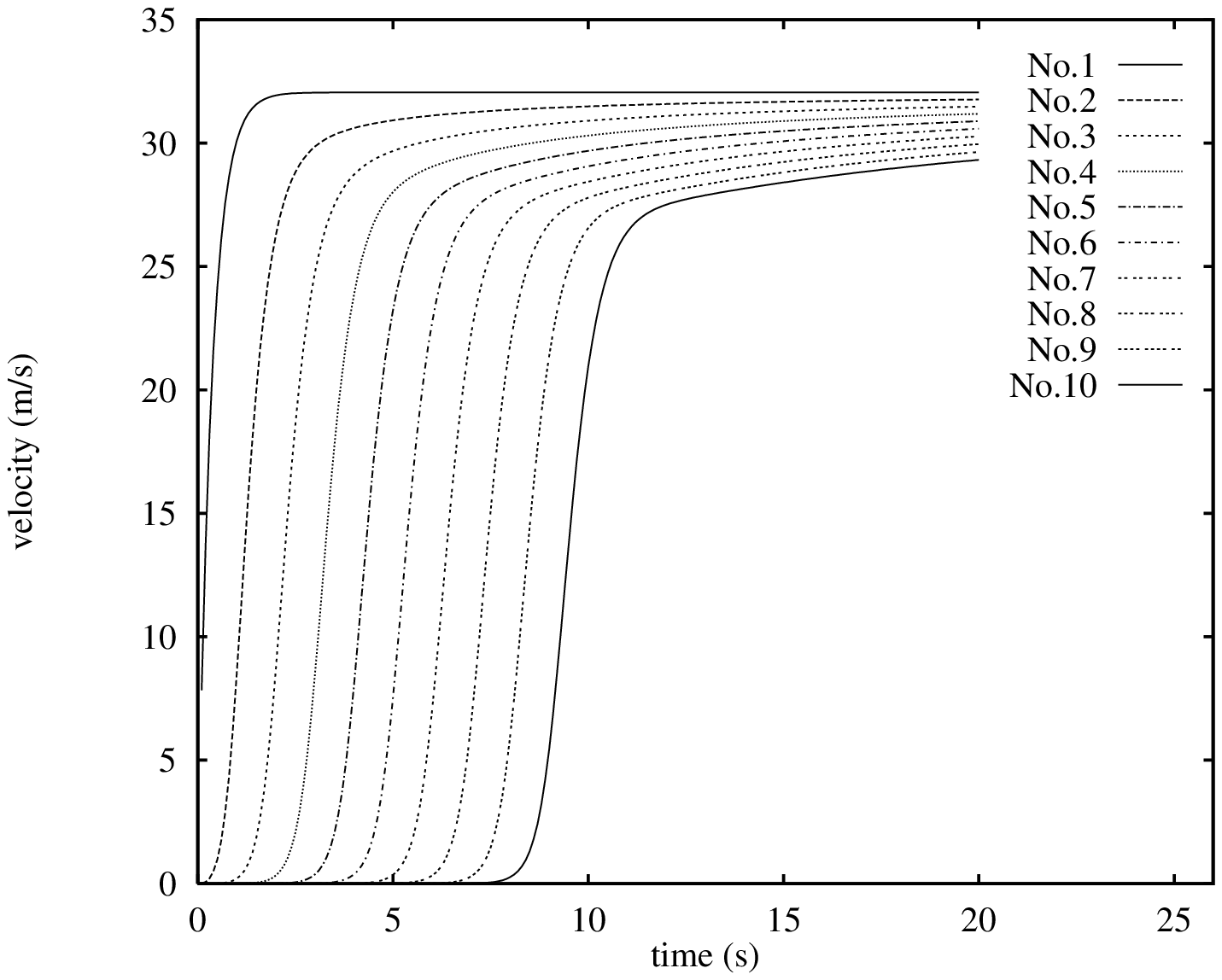}
\epsfxsize=7.5cm
\epsfbox{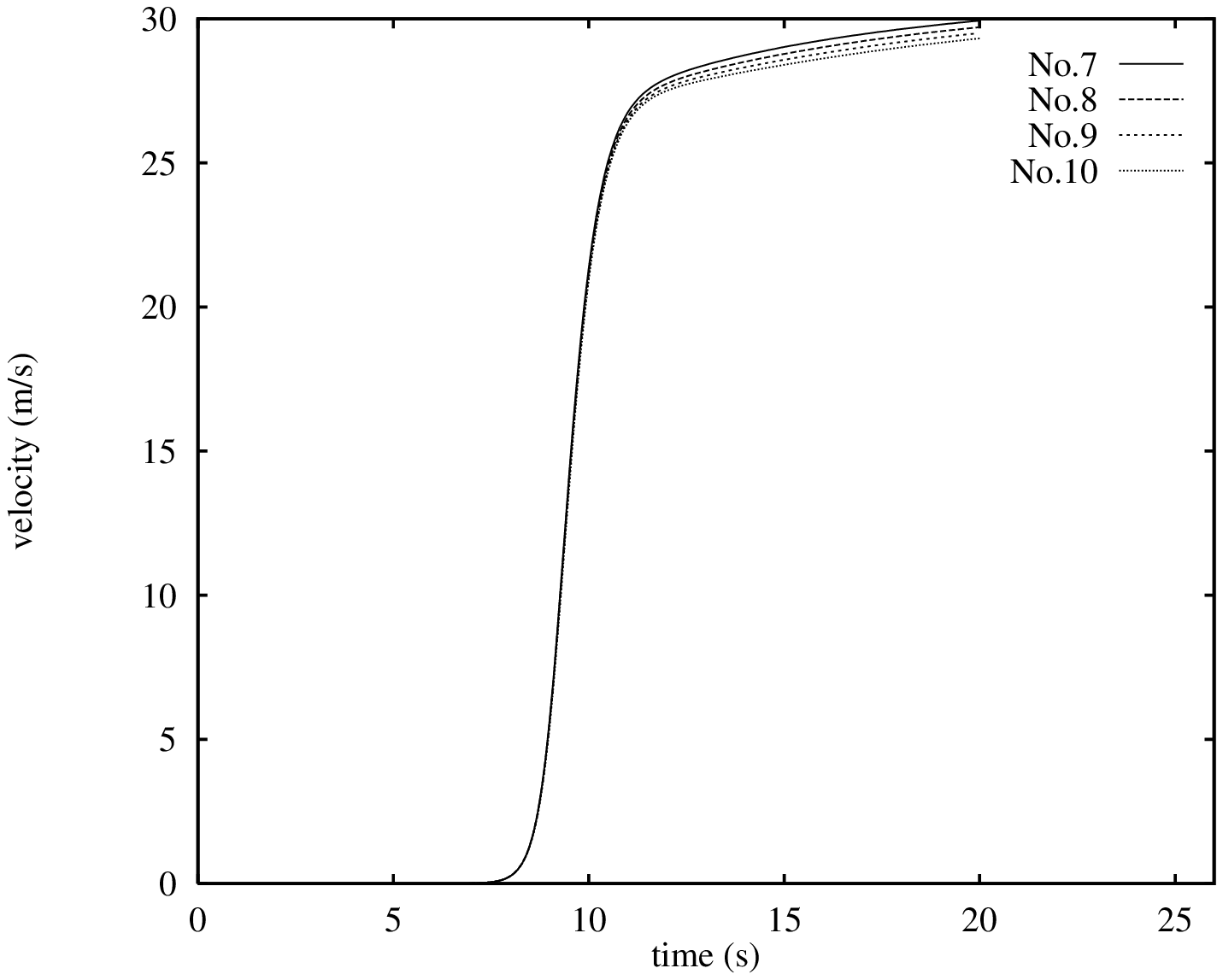}
\caption{(a) Behavior of velocities of first ten vehicles under
traffic lights with $a=2.8~{\rm s}^{-1}$.
(b) Figure of shifted curves $\dot x_n(t-(n-10)T)$ with $T=1.03$~s.}
\label{signal.2.8}
\end{figure}

We perform the numerical simulation for the cases $a=2.0,\ 2.8\ {\rm
s}^{-1}$ and obtain the time dependence of velocity $\dot x_n(t)$ of
$n$-th vehicles in a queue. Figures \ref{signal.2.0}(a) and \ref
{signal.2.8}(a) show the behavior of motion for the first ten
vehicles. In the figures, it seems that the behaviors of motion for
vehicles on the down stream converge into a common shape. Thus, except
first several vehicles, every vehicle in the queue almost replicates
the behavior of its preceding one with a certain delay time. In this
case, the relation (\ref{eq:def_delay}) can be applied 
for n-th vehicle with enough large number $n$. In this sense 
we may say that replicative pattern of vehicle motion 
is realized ``asymptotically'' and only in this occasion 
we can define a delay time of motion $T$ in just the same 
way as we defined in section 2.  Note that
this definition of the delay time slightly differs from a time lag
with which the successive vehicles start for example.

From Figures \ref{signal.2.0}(a) and \ref{signal.2.8}(a), we can
estimate the delay time $T = 1.10$~s for $a=2.0~{\rm s}^{-1}$ and $T
= 1.03$~s for $a=2.8~{\rm s}^{-1}$.  To confirm this similarity of
vehicle motion, we also show plots of translated data $\dot
x_n(t-(n-10)T)$ for the 7th, 8th, 9th and 10th vehicles in Figures \ref
{signal.2.0}(b) and \ref{signal.2.8}(b).

\section{Delay Time in Highway Traffic Flow}
Next we investigate delay time under a simple situation where $N$
vehicles move on a single lane circuit with circumference $L$. Of
course we assume that road conditions are uniform along the circuit
and drivers are identical. Numerical calculations are made with the
initial condition: $x_n = 2\delta_{n1}- b(n-1) \ ( 1\le n\le N)$,
$\dot x_n=V(b)$. The first term of the right hand side of this
condition means that a small perturbation is added to the first
vehicle $x_1$.

In the previous papers \cite{AichiA}, we have shown a homogeneous flow
changes into congested flow spontaneously if the density of vehicles
is greater than the critical value. The results of simulations
indicate that after enough time the traffic flow on a circuit
creates an alternating pattern of high and low density regions. The
motion of vehicles in this flow is visualized by plotting them 
in the `phase space'
($\Delta x$, $\dot{x}$). After the traffic flow becomes stationary,
the trajectory of every vehicle in this `phase space' draws a kind of
limit cycle which we named `hysteresis loop' in Ref.\cite{AichiA} (see
also Fig. \ref{limit cycle}).

Now let us estimate delay times for two cases under this traffic flow: (A)
the first stage and (B) the final stationary-state stage.
\begin{figure}[htb]
\hspace*{-0.3cm}
\epsfxsize=7cm
\epsfbox{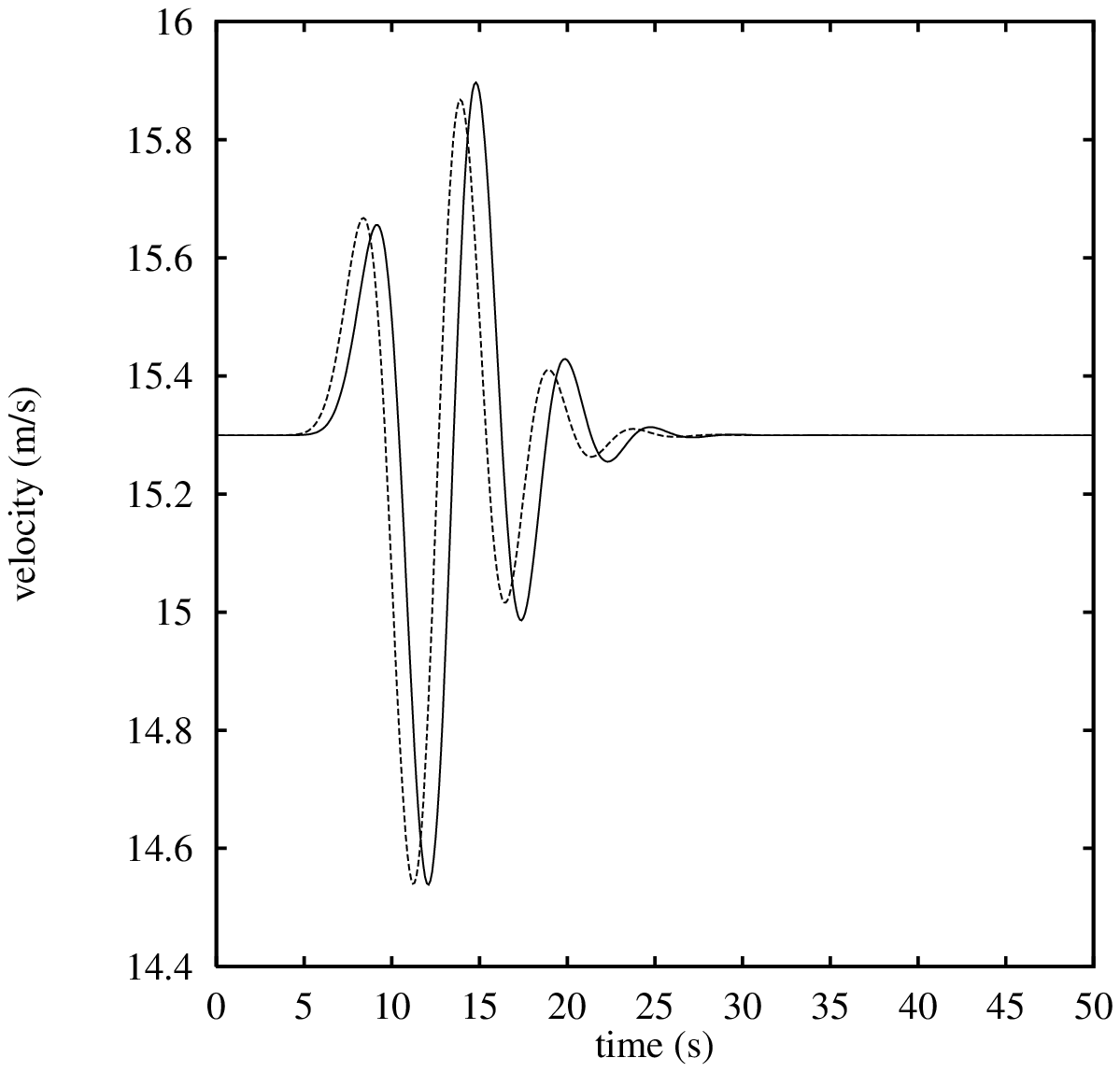}
\epsfxsize=7cm
\epsfbox{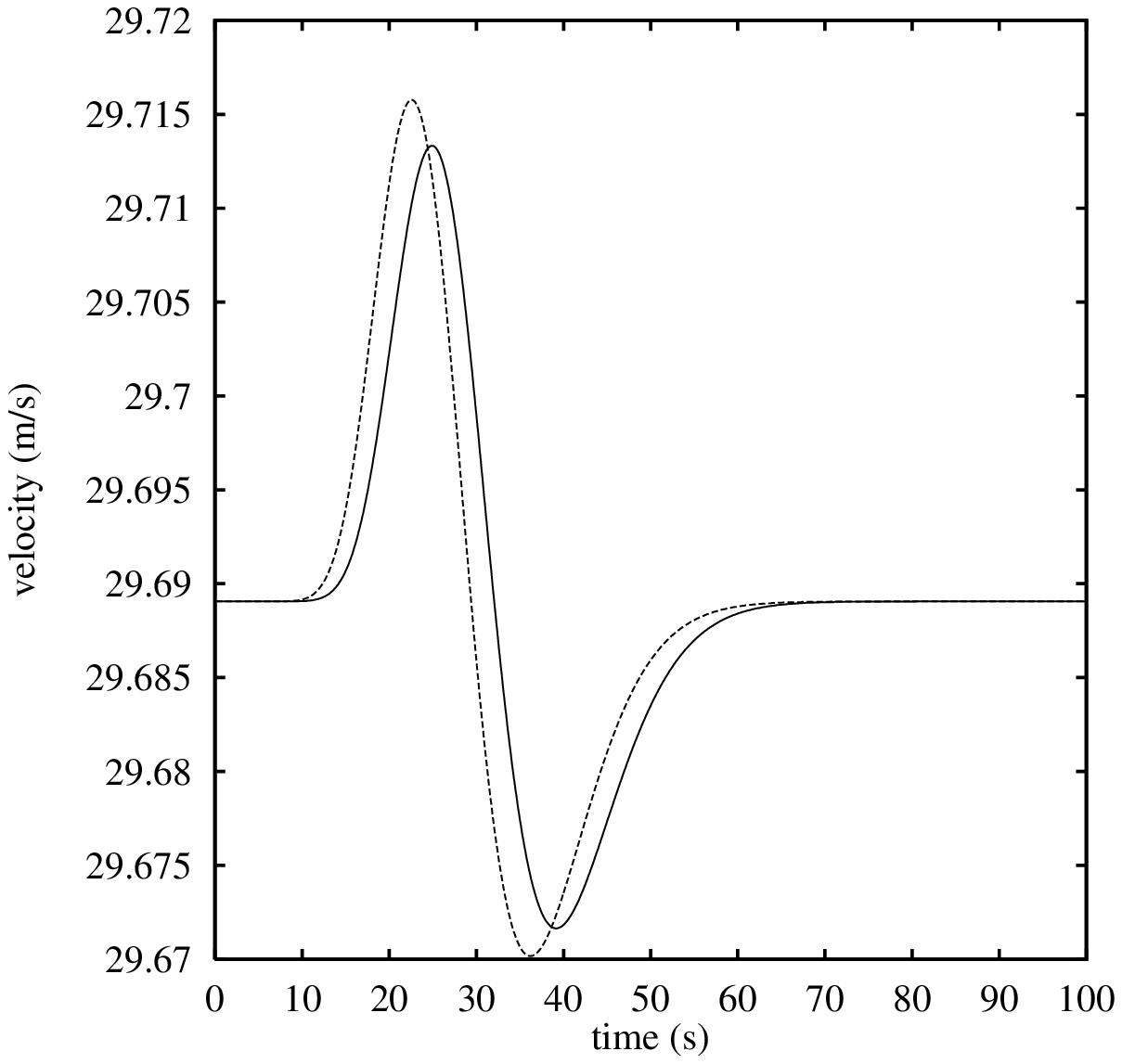}
\caption{Motions of 10th and 11th vehicles in the first stage:
(a) $\Delta x=$25~m for 50 seconds and (b) $\Delta x=$40~m for 100 seconds
with $a=2.0\ {\rm s}^{-1}$.}
\label{fig:highway.caseA}
\end{figure}

\noindent
$\cdot$ Case A

 In this case the traffic flow is almost homogeneous. Let us
pick up a pair of vehicles $n=10,11$. A small perturbation of the
first vehicle $x_1$ propagates backward and after several seconds
those pair of vehicles change their velocities. The typical behaviors 
are demonstrated in Figures \ref{fig:highway.caseA}.
In the same way as section 2, the delay
time of motion can be estimated from numerical results.
The obtained values of delay times are shown in Table
\ref{tab:highway.delay2.0} for the case of $a=2.0~{\rm s}^{-1}$ 
and Table
\ref{tab:highway.delay2.8} for the case of $a=2.8~{\rm s}^{-1}$. 
As references, the delay times for low frequency limit and 
for the enhanced mode $\omega_0$ are also shown 
in Tables \ref{tab:highway.delay2.0}, \ref{tab:highway.delay2.8}.
From these results, the low frequency limit is a good approximation
and the delay time $T$ is almost independent on sensitivity $a$ for
stable traffic flow. For unstable case, there exist contributions from
blow-up modes and they have sensitivity dependence.
\begin{table}[hbt]
\begin{center}
\caption{Delay times for various headway with $a=2.0\ {\rm s}^{-1}$.
The second column indicates that traffic flow is stable (-) or unstable
(+). The third and fourth columns show analytical results given in
section 2.}
\label{tab:highway.delay2.0}
\vspace{0.3cm}
\begin{tabular}{|c|c|c|c|c|}\hline
$\Delta x$ (m) & $f-a/2$ & $T_0=f^{-1}$ (s) 
& $T_{\rm enhanced}$ (s) & $T_{\rm simulation}$ (s) \\ \hline
10 & $-$ & 2.6427 &    -   & 2.6 \\ \hline
15 & $-$ & 1.3434 &    -   & 1.35 \\ \hline
20 & $+$ & 0.8282 & 0.8884 & 0.95 \\ \hline
25 & $+$ & 0.6921 & 0.8017 & 0.85 \\ \hline
30 & $+$ & 0.8282 & 0.8884 & 0.95 \\ \hline
35 & $-$ & 1.3434 &    -   & 1.35 \\ \hline
40 & $-$ & 2.6427 &    -   & 2.6 \\ \hline
50 & $-$ & 13.101 &    -   & 13 \\ \hline
\end{tabular}
\caption{Delay times for various headway with $a=2.8\ {\rm s}^{-1}$.
The second column indicates that traffic flow is stable (-) or unstable
(+). The third and fourth columns show analytical results given in
section 2.}
\label{tab:highway.delay2.8}
\vspace{0.3cm}
\begin{tabular}{|c|c|c|c|c|}\hline
$\Delta x$ (m) & $f-a/2$ & $T_0=f^{-1}$ (s)
& $T_{\rm enhanced}$ (s) & $T_{\rm simulation}$ (s)\\ \hline
10 & $-$ & 2.6427 &    -   & 2.6 \\ \hline
15 & $-$ & 1.3434 &    -   & 1.35 \\ \hline
20 & $-$ & 0.8282 &    -   & 0.85 \\ \hline
25 & $+$ & 0.6921 & 0.6996 & 0.75 \\ \hline
30 & $-$ & 0.8282 &    -   & 0.85 \\ \hline
35 & $-$ & 1.3434 &    -   & 1.35 \\ \hline
40 & $-$ & 2.6427 &    -   & 2.6 \\ \hline
50 & $-$ & 13.101 &    -   & 13 \\ \hline
\end{tabular}
\end{center}
\vspace*{-0.4cm}
\end{table}

\noindent
$\cdot$ Case B

\begin{figure}[htb]
\hspace*{3cm}
\epsfxsize=7cm
\epsfbox{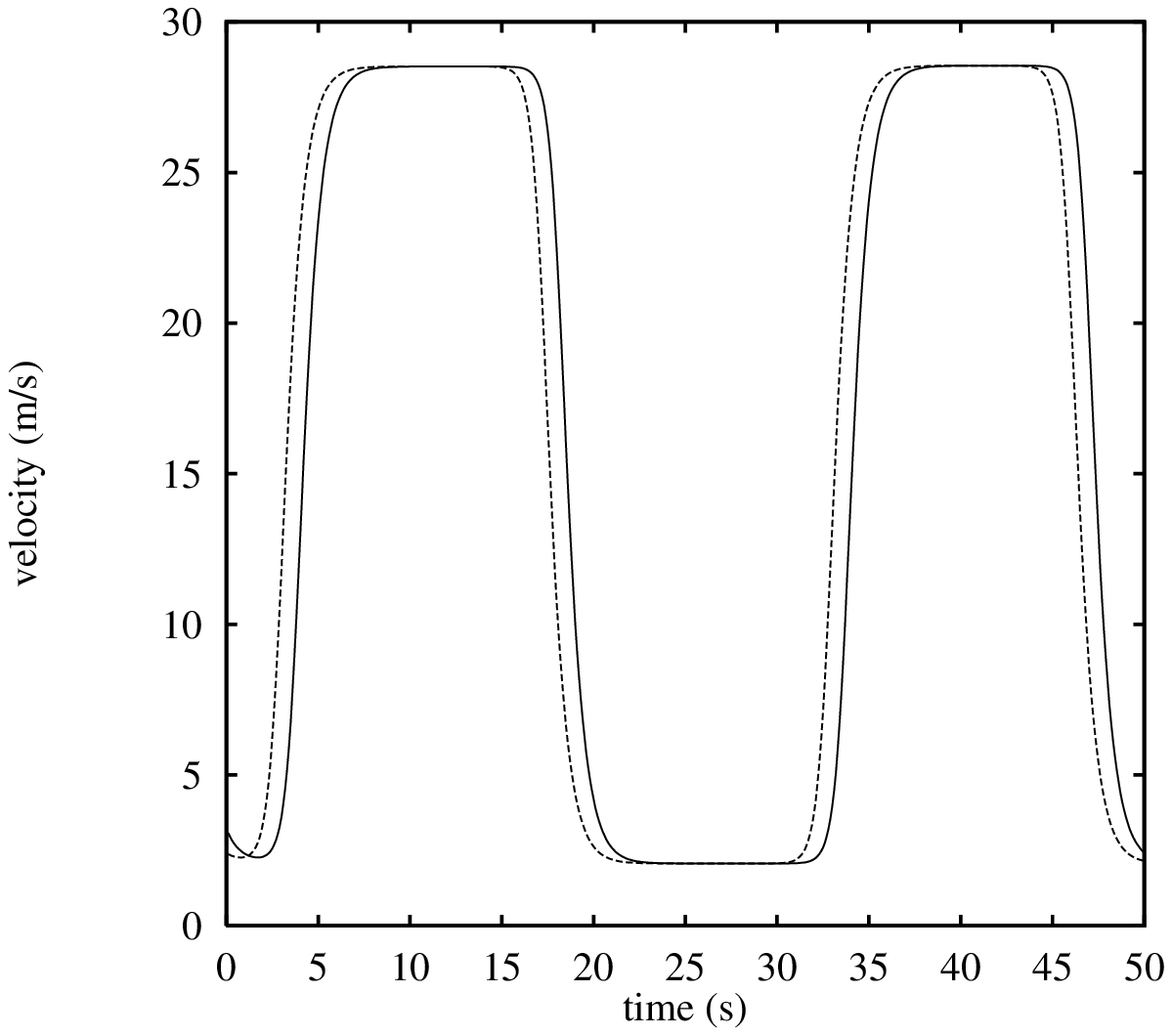}
\caption{Motion of successive two vehicles in congested flow.
Initial condition is $\Delta x=$25~m with $a=2.0~{\rm s}^{-1}$.}
\label{fig:highway.caseB}
\end{figure}

\begin{figure}[hbt]
\hspace*{3cm}
\epsfxsize=8cm
\epsfbox{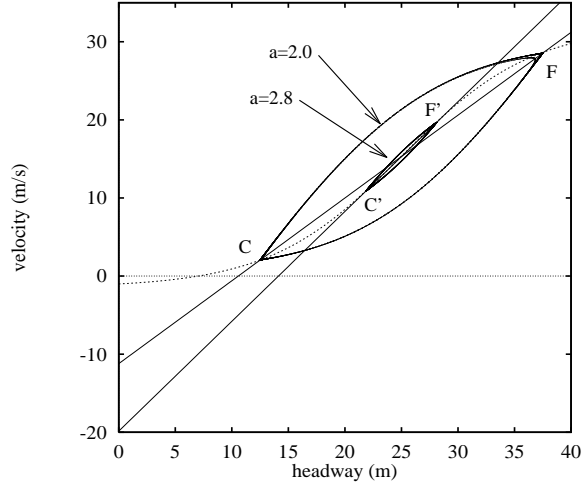}
\caption{The limit cycles for $a=2.0~{\rm s}^{-1}$ and $a=2.8~{\rm
s}^{-1}$.}
\label{limit cycle}
\end{figure}

After sufficiently long time, traffic flow forms
stationary patterns of high and low density regions.
Under this situation vehicle does not change its velocity unless it
encounters a boundary of high and low density regions. Let us observe
the motion of a vehicle on the boundary. A vehicle which encounters
the boundary changes its velocity. After some delay time, the following vehicle
comes to the boundary and changes its velocity in the same way as
the previous vehicle. The typical behavior of vehicles is shown in
Figure \ref{fig:highway.caseB}.

The delay time of vehicle motion in this case can be derived as follows.
Consider two vehicles: one enters into a congested
region from a free region and
after a certain interval $T$, the next one follows.
In the `phase space' ($\Delta x$, $\dot{x}$) (Fig.\ref{limit cycle}), 
the free region is denoted as
a point F$(\Delta x_F,v_F)$, that is, vehicles is moving with
velocity $v_F$ and headway $\Delta x_F$.
The delay time $T$ is defined as the time which is needed for the next 
vehicle to reach the boundary and enter into a congested region. 
Now at the time when the first vehicle reaches the boundary of the free region 
the distance between the next vehicle and this boundary is of course 
$\Delta x_F$. The next vehicle runs with velocity $v_F$ and the
boundary itself also is moving backward with velocity $v_B$, 
so if the vehicle and the boundary will meet after time interval $T$,
we can write the following relation
\begin{equation}
      v_FT+v_BT=\Delta x_F\ .
\label{eq:Tf}
\end{equation}
Similar relation can be written at a boundary where a vehicle exits
from congested region. 
\begin{equation}
    v_CT+v_BT=\Delta x_C\ . 
\label{eq:Tc}
\end{equation}
This can be confirmed if one recalls that the pattern of the flow
is already stationary, and the input vehicles of a boundary of
congested region must be equal to the output of another boundary. 
The above two equations,(\ref{eq:Tf}) and (\ref{eq:Tc}) indicate that the 
time interval $T$ and $v_B$ can be graphically expressed as the slope
and the intercept with vertical axis of the line connecting F and C for
$a=2.0\ {\rm s}^{-1}$ and F' and C' for $a=2.8\ {\rm s}^{-1}$ in Figure
\ref{limit cycle}.  If we write the velocity of the first vehicle as
$v(t)$, then the one of the following vehicle is $v(t-T)$, which
implies that $T$ is just the delay time of vehicle-motion defined in
the previous section.

It may be convenient to adopt the coordinate moving with the
congestion pattern. Let $x(t)$ and $X(t)$ be the positions of a
vehicle measured in a fixed and a moving (with constant velocity
$v_B$) coordinates, respectively;
\begin{equation}
x(t) = X(t) + v_Bt.
\label{eq:y2x}
\end{equation} 
Then we get
\begin{eqnarray}
 \ddot{X}(t)
 &=& a\{ V(\Delta X(t))-(\dot{X}(t)+v_B)\} \nonumber \\
 &=& a\{ W(\Delta X(t))-\dot{X}(t)\},
\label{eq:E33}
\end{eqnarray}
where $W(\Delta x)$ is the optimal velocity function in the co-moving
frame;
\begin{equation}
W(\Delta x) = V(\Delta x) - v_B.
\label{eq:E34}
\end{equation}

Let us concentrate on the motion of vehicles in this co-moving
coordinate. Every vehicle passes a same position with a certain time
intervals $T$; in high density region, the time interval is given by
\begin{equation}
T_C = \left(\frac{\Delta X}{\dot X}\right)_C,
\end{equation}
and (similarly) on the point $F$, we have,
\begin{equation}
T_F = \left(\frac{\Delta X}{\dot X}\right)_F.
\end{equation}
These two time intervals should be equal since otherwise the
congestion pattern moves. We write this time interval by $T$;
\begin{equation}
T = T_C = T_F,
\end{equation}
which is of course equivalent to Eqs.(\ref{eq:Tf}) and (\ref{eq:Tc}).

We have carried out numerical simulations. For sensitivity $a=2.0~{\rm
s}^{-1}$, we find $C(12.51,2.05)$ and $F(37.50,28.55)$ yielding the
delay of vehicle motion $T=0.943$~s and the back velocity of the
boundary $v_B=11.2$~m/s. As for $a=2.8~{\rm s}^{-1}$,
$C'(21.89,10.92)$ and $F'(28.11,19.68)$ yielding $T=0.711$~s and
$v_B=19.9$~m/s.

It is interesting to find that the main contribution of the resultant
delay of vehicle-motion comes from the structure of the Optimal
Velocity Model and not
from the explicit delay $\tau$.

\section{Summary and Discussions}
The notion of delay time of response $\tau$ has played a significant
role in the history of traffic dynamics. Indeed delays of vehicle
motions are observed in many cases, in traffic lights waiting queues
or in highway traffic motions and the delay time are usually observed
to be of order 1 second. We should take account of the effect of
observed delay time
and it has long been thought that it must be introduced in the
equation of motion as an explicit delay time, most of which is caused
by driver's physiological delay time and mechanical delay of response
of vehicles. However it is known that the physiological response time
is of order 0.1 second, not of order 1 second. We should be careful
that the delay time of vehicle motion comes from another
origin, that is, from the equation of motion itself
which we have here investigated intensively. The results are
summarized as follows:

\begin{enumerate}
\item{ The case of  a leader vehicle and its follower}\\
As is seen in Figure \ref{fig:eta-delay}(b), if the headway distance is
around 25~m in which drivers are sensitive to the behavior of the
motion of the preceding vehicle, we clearly recognize delay time is
around 1 second independently of the frequency of the leader's
velocity-change function $\lambda(t)$. However if their headway
distance is more than 40~m, delay time is estimated to be larger than
1 second, and sometimes we obtain 6 second for the case $\Delta x=6$~m
for the case of low frequency ($\omega \sim 0$). 
This is because of the structure of
optical velocity function. If the slope of the optimal velocity
function is very
small, drivers are insensitive to the behavior of the preceding
vehicle. This can be easily understood if one considers the extreme
case in which the function $V$ is independent of $\Delta x$ (and so
$f=0$). In this case, a follower never reacts to its previous vehicle and
accordingly its delay time of motion becomes infinity.
\item{ A queue of vehicles controlled by traffic lights}\\
In this case except the first several vehicles, most of the
succeeding vehicles behave almost similarly as seen in Figures \ref
{signal.2.0} and \ref {signal.2.8}. From those figures,
delay times are read off; $T=1.10$~s for $a=2.0~{\rm s}^{-1}$ and
$T=1.03$~s for $a=2.0~{\rm s}^{-1}$. Although $T$ depends on the
sensitivity adopted, the results obtained is again of order 1 second
for reasonable realistic sensitivity.
\item{Many-vehicle case of highway traffic flow}\\
In Figures \ref{fig:highway.caseA} we show the typical behaviors of a
pair of vehicles and in Table \ref{tab:highway.delay2.0} and \ref
{tab:highway.delay2.8} delay times obtained by numerical simulations
are summarized with various values of its headway $\Delta x$. Again in
the case of $\Delta x =25$~m, the estimated delay time is found to be
of order 1 second, and for the larger $\Delta x$, the larger the value
of delay time is obtained. In the case where the congested flow
becomes stable, $T=0.71~{\rm s}$ for $a=2.8~{\rm s}^{-1}$ and
$T=0.9471~{\rm s}$ for $a=2.0~{\rm s}^{-1}$. Since the end point C (F) on
the limit cycle becomes larger (smaller) as $a$ becomes larger (see
for example Fig.7 in ref.\cite{AichiC}), so of course $T$ becomes
smaller for larger $a$ (high sensitivity).
\end{enumerate}

All our results
show that the estimated delay time in our OVM is almost enough to
reproduce the order of observed delay time. It now becomes obvious that the 
delay time of motion arises as an effect of dynamical 
equation itself without any explicit introduction of $\tau$.
It is interesting to find the main contribution of the resultant delay
of vehicle motion are of same order of magnitudes in any cases. 
This may come from the structure of optical velocity
function itself which we have determined phenomenologically. 
However we believe that this remarkable fact has 
its more profound reason, which  will be made clearer by further 
investigation on structure of OVM by performing analytical study.

\end{document}